\documentclass{PoS}

\usepackage{epsfig}

\def\beq{\begin{equation}}
\def\eeq{\end{equation}}
\def\beqa{\begin{eqnarray}}
\def\eeqa{\end{eqnarray}}

\title{Single top quark production cross section at hadron colliders}
 
\ShortTitle{Single top quark production cross section at hadron colliders}

\author{\speaker{Nikolaos Kidonakis}%
        \thanks{This work was supported by the National Science 
Foundation under Grant No. PHY 0855421.}\\
       Kennesaw State University, USA\\
       E-mail: \email{nkidonak@kennesaw.edu}}

\abstract{
I present results for the theoretical cross section of single top
quark production in the $s$ channel and also via associated production with a 
W boson or a charged Higgs. The calculations include higher-order corrections 
at NNLL accuracy. Detailed numerical results are provided for Tevatron and LHC 
energies.
}

\FullConference{XVIII International Workshop on Deep-Inelastic Scattering and Related Subjects\\
                 April 19 -23, 2010\\
                 Convitto della Calza, Firenze, Italy}

\begin{document}

\section{Introduction}

Single top quark production has been obeserved at the Tevatron 
\cite{D0st,CDFst} with a cross section consistent with Standard Model 
predictions. There are three partonic channels: the $t$-channel, 
the $s$-channel, and associated $tW^-$ production. A related process 
to $tW^-$ production is associated production of a top quark with a 
charged Higgs boson, $bg \rightarrow tH^-$.

The NLO corrections for the $s$ and $t$ channels were given in 
\cite{bwhl}, for $tW^-$ production in \cite{ZhutW}, and for
$tH^-$ production in \cite{ZhutH,Plehn}.
Higher-order QCD corrections are significant for single top quark production.
Soft-gluon corrections resummed at next-to-leading-logarithm (NLL) 
accuracy have been calculated for the three single top  
channels in \cite{NKst,NKst2} and for $tH^-$ in 
\cite{NKchiggs,NKchiggs2}. 

Recent developments in two-loop calculations of soft anomalous dimensions 
\cite{NK2l,NKs} have now made it possible to improve the resummation to 
next-to-next-to-leading-logarithm (NNLL) accuracy.
In this contribution I discuss new approximate NNLO calculations, derived 
from the expansion \cite{NKNNNLO} of the NNLL resummed cross section, 
for $s$-channel production at the Tevatron and the LHC 
and for associated production of a top with a $W^-$ 
or with a charged Higgs \cite{NKs}.

\section{Single top quark production in the $s$ channel}

The $s$-channel processes in single top quark production are 
$q{\bar q}' \rightarrow {\bar b} t$ 
(i.e. $u {\bar d} \rightarrow {\bar b} t$, etc).
We calculate the cross section at Tevatron and LHC energies using the 
MSTW2008 NNLO pdf \cite{MSTW2008}.

At NLO the soft gluon corrections dominate the cross section and provide 
an excellent approximation to the complete NLO corrections, especially 
when damping factors are used to limit contributions far from threshold
\cite{NKs} (as are also used for $t{\bar t}$ production \cite{NKRV}).    
Thus it can be reasonably expected that at NNLO the soft-gluon corrections 
are dominant and that the NNLO cross section derived from NNLL resummation 
is a good approximation to the complete result.

\begin{figure}
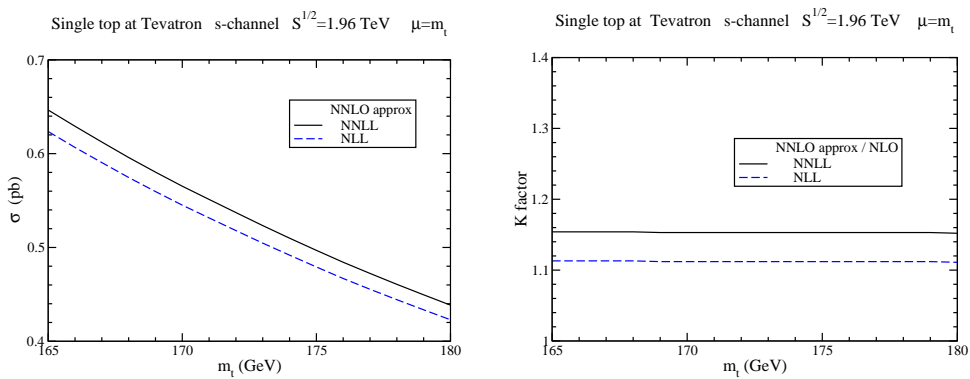

\begin{center}
\includegraphics[width=0.4\textwidth]{tevschmtplot.eps}
\hspace{5mm}
\includegraphics[width=0.4\textwidth]{Ktevschmtplot.eps}
\caption{Cross section (left) and $K$ factor (right) for $s$-channel single top quark production 
at the Tevatron.}
\end{center}
\end{figure}

In Fig. 1 we plot the cross section and the $K$ factor for $s$-channel 
single top quark production at the Tevatron. Results are shown for the NNLO 
approximate cross section from both NLL and NNLL resummation. The improved 
NNLL result increases the cross section even more than NLL, and provides 
a 15\% enhancement over the NLO cross section. 
For a top quark mass of 173 GeV we find 
\beqa
\sigma^{\rm NNLOapprox,\, top}_{s-{\rm channel}}(m_t=173 \, {\rm GeV},  \, 1.96 \, {\rm TeV})&=&0.523 {}^{+0.001}_{-0.005} {}^{+0.030}_{-0.028} \; {\rm pb} 
\nonumber
\eeqa
where the first uncertainty is from scale variation and the second from 
pdf errors at 90\% C.L.
We note that the cross section for $s$-channel single antitop production at 
the Tevatron is identical to that for single top.

\begin{figure}
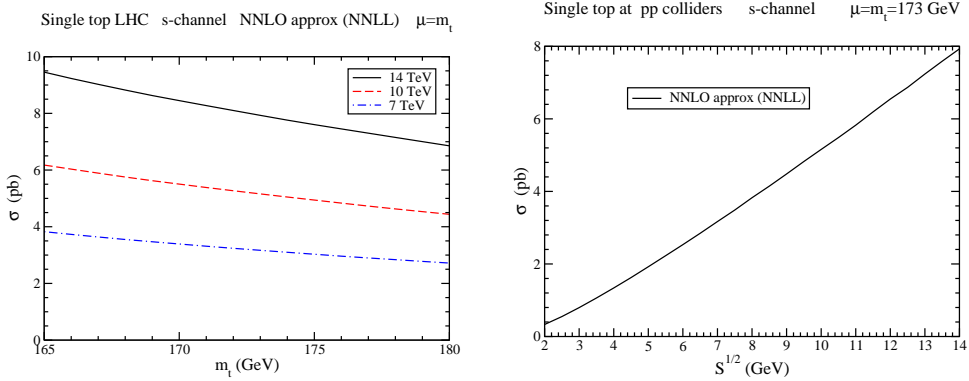

\begin{center}
\includegraphics[width=0.4\textwidth]{lhctopschmtplot.eps}
\hspace{5mm}
\includegraphics[width=0.4\textwidth]{lhcscapschmtplot.eps}
\caption{Cross section for $s$-channel single top quark production at the LHC 
versus mass (left) and energy (right).}
\end{center}
\end{figure}

In Fig. 2 we plot the cross section for $s$-channel 
single top quark production at LHC energies. 
The left plot shows the NNLO approximate cross section from NNLL resummation
versus the top quark mass for the current LHC energy of 7 TeV as well as for  
energies of 10 TeV and 14 TeV. The enhancement over NLO is around 13\% at 
all three energies. The plot on the right shows the cross section as a 
function of c.m. energy at the LHC. 
For a top quark mass of 173 GeV we find at 7 TeV 
\beqa
\sigma^{\rm NNLOapprox,\, top}_{s-{\rm channel}}(m_t=173\, {\rm GeV}, \, 7\, {\rm TeV})&=&3.17 \pm 0.06 {}^{+0.13}_{-0.10} \; {\rm pb}
\nonumber
\eeqa
while at 10 TeV we have  $5.16 \pm 0.09 {}^{+0.20}_{-0.14}$ pb, 
and at 14 TeV we have $7.93 \pm 0.14 {}^{+0.31}_{-0.28}$ pb.

\begin{figure}
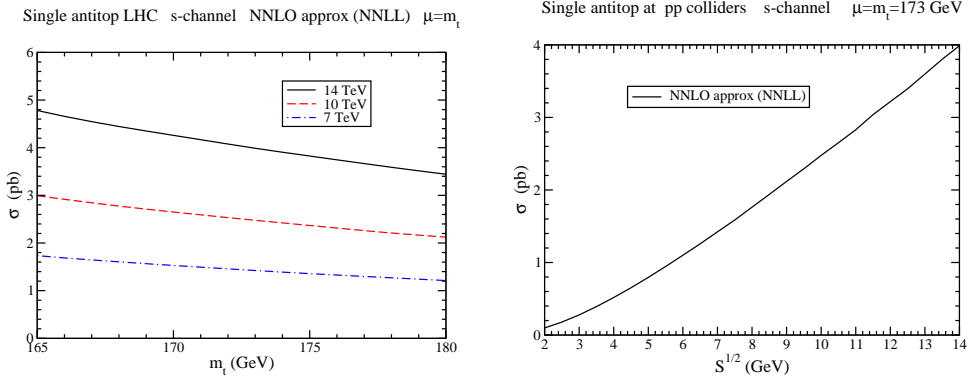

\begin{center}
\includegraphics[width=0.4\textwidth]{lhcatopschmtplot.eps}
\hspace{5mm}
\includegraphics[width=0.4\textwidth]{lhcatscapschmtplot.eps}
\caption{Cross section for $s$-channel single antitop  production at the LHC 
vs mass (left) and energy (right).}
\end{center}
\end{figure}

At the LHC the antitop $s$-channel cross section is different than for top. 
Figure 3 shows the corresponding results.
At 7 TeV we find
\beqa
\sigma^{\rm NNLOapprox,\, antitop}_{s-{\rm channel}}(m_t=173\, {\rm GeV}, \, 7\, {\rm TeV})&=&1.42 \pm 0.01 {}^{+0.06}_{-0.07} \; {\rm pb}
\nonumber 
\eeqa
while at 10 TeV we have $2.48 \pm 0.02 {}^{+0.09}_{-0.13}$ pb, 
and at 14 TeV we have $3.99 \pm 0.05 {}^{+0.14}_{-0.21}$ pb.

\section{Associated production of a top quark with a $W^-$}

\begin{figure}
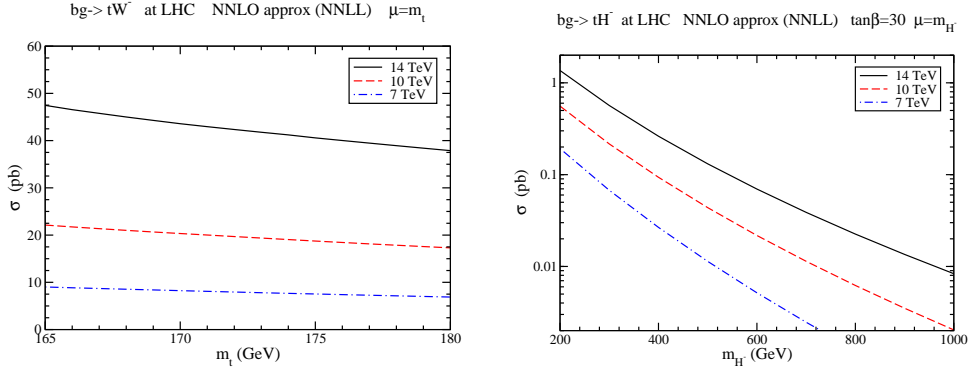

\begin{center}
\includegraphics[width=0.4\textwidth]{lhctWmtplot.eps}
\hspace{5mm}
\includegraphics[width=0.4\textwidth]{chiggsplot.eps}
\caption{Cross section for $tW^-$ (left) and $tH^-$ (right) production at the LHC.}
\end{center}
\end{figure}

We now turn to the process $bg \rightarrow tW^-$. This process is 
negligible at the Tevatron but significant at the LHC.
The left plot of Fig. 4 shows the approximate NNLO cross section from NNLL 
resummation for $tW^-$ production at LHC energies. The enhancement over NLO 
is around 8\%.
At 7 TeV with $m_t=173$ GeV we find 
\beqa
\hspace{2cm} \sigma^{\rm NNLOapprox}_{tW^-}(m_t=173 \, {\rm GeV}, \, 7\, {\rm TeV})&=&7.8 \pm 0.2 {}^{+0.5}_{-0.6} \; {\rm pb}
\nonumber
\eeqa
while at 10 TeV we have $19.4 \pm 0.5 {}^{+1.0}_{-1.1}$ pb, 
and at 14 TeV we have $41.8 \pm 1.0 {}^{+1.5}_{-2.4}$ pb.
Again, the first uncertainty is from scale variation and the second is from 
the MSTW2008 NNLO pdf at 90\% C.L.

The cross section for ${\bar t}W^+$ production is identical to that 
for $tW^-$.

\section{Associated production of a top quark with a charged Higgs}

Finally we consider the process $bg \rightarrow tH^-$. Charged Higgs bosons 
appear in supersymmetric theories, and the associated production of a top quark
with an $H^-$ is a potentially relevant process. The right plot of Fig. 4 
shows the approximate NNLO cross section from NNLL resummation for $tH^-$ 
production at LHC energies, for a value of $\tan\beta=30$, versus the 
charged Higgs mass. 
The NNLO approximate corrections increase the NLO cross section by 
$\sim 15$\% to $\sim 20$\% for the range of charged Higgs masses shown.

\end{document}